\documentclass{article}
\usepackage{amsmath}
\usepackage{amssymb}
\usepackage{graphicx}
\usepackage{epstopdf}

\def\beq{\begin{equation}}
\def\eeq{\end{equation}}
\def\bea{\begin{eqnarray}}
\def\eea{\end{eqnarray}}
\def\ar{\begin{array}}
\def\ear{\end{array}}

\def\nn{\nonumber}

\def\Om{\Omega}
\def\ka{\kappa}

\def\la{\lambda}

\def\rd{{\rm d}}

\def\Ra{{\Rightarrow}}

\def\rd{{\rm d}}

\begin{document}
\begin{center}
{\Large Conformal transformations and the beginning of the Universe}\\[5mm]

Krzysztof A. Meissner$^{1}$ and Pawe\l~ Nurowski$^{2}$\\[3mm]
{\it $^1$ Faculty of Physics,
University of Warsaw,\\
Pasteura 5, 02-093 Warsaw, Poland\\
$^2$ Center for Theoretical Physics of the Polish Academy of Sciences\\
Al. Lotnik{\'o}w 32/46, 02-688 Warsaw, Poland
}

\begin{abstract}
\noindent We show that two consecutive Universes with positive cosmological constants filled with perfect fluids and conformal to the positive-spatial-curvature FLRW metric by an analytic conformal transformation have the following features implied by the Einstein equations: a) the fluids can only belong to 4 classes (radiation, dust and two other classes with negative pressures corresponding to a gas of strings and a gas of domain walls), b) the field equations on both sides of the future/eternity hypersurface exhibit certain duality with the Einstein Universe at the background, and c) both eons (one at the end, and the other at the beginning) are almost critical, so that the future eon is dominated by radiation and resembles the beginning of our Universe.
\end{abstract}
\end{center}

\vspace{0.2cm}

\noindent
In Penrose's Conformal Cyclic Cosmology  \cite{P} the metric $\check{g}$ of the Universe is conformally flat at the surface $t=0$ of the initial singularity \cite{PT1}. Penrose extends a \emph{regular} representative $g$ of the conformal class of $\check{g}$ from $t=0$ to a \emph{regular} metric $g$ in a strip given by $t\in ]-\epsilon, \epsilon]$, calls it \emph{intermediate metric}, and relates $g$  in the `around the Big Bang' region $t\in ]-\epsilon,\epsilon[$ -- the \emph{`bandage region'}, as he calls it -- to two \emph{physical} metrics:
    \begin{itemize}
      \item the metric $\check{g}$ describing the Universe close to the singularity, when $t\in]0,\epsilon[$, and
      \item the metric $\hat{g}$ which, is interpreted as the physical metric of the previous Universe (previous eon), when $t\in]-\epsilon,0[$.\end{itemize}
Formally, having chosen the intermediate metric $g$, one gets three metrics: $\hat{g}$, $g$  and $\check{g}$ in the entire bandage region $t\in]-\epsilon,\epsilon[$. They are related via:
\beq
\hat{g}=\frac{1}{f^2}\,g,\ \ \ \ \check{g}=f^2 g
\label{ghcmetric}
\eeq
where $f=f(t)$, and $f(t)$ is chosen in such a way that $\check{g}$ coincides with the metric $\check{g}$ of the current Universe (current eon), when $t\in [0,\epsilon[$, and $\hat{g}$ coincides with the physical metric $\hat{g}$ of the previous eon, when $t\in]-\epsilon, 0]$.

In this brief note we provide a simple model in which the Einstein equations for the metrics $\hat{g}$ and $\check{g}$ can be consistently solved, when one assumes that the two consecutive eons have positive cosmological constants and are filled with perfect fluids. A framework for the model was discussed in \cite{PT2,ETN}.

In our model the intermediate metric $g$ is a Friedman-Lema{\^i}tre-Robertson-Walker (FLRW) metric
\beq
g=\left[-\rd t^2+\frac{h^2}{\left(1+\frac{\ka}{4}(x^2+y^2 +z^2)\right)^2}(\rd x^2+\rd y^2 +\rd z^2)\right]
\label{gmetric}
\eeq
with spatial curvature $\ka={\rm const}$ and with $h=h(t)$.

We assume that the metric $g$ is regular at $t=0$, so that the singularity of $\check{g}$ at the transition ($t\to 0$) is due to the behaviour of the conformal factor $f(t)\to 0$ in $\check{g}$. We therefore have: $f(t)=Ct+O(t^2)$ and $h(t)=1+O(t)$ where the linear dependence of $f(t)$ on $t$ for $t\to 0$ comes from the assumption of the dominance of the cosmological constant (de Sitter solution) at the end of the previous eon.

The dynamics of the model is governed by the Einstein equations, satisfied by each of the metrics $\hat{g}$ and $\check{g}$ separately. The Einstein equations, respectively for $\hat{g}$ and $\check{g}$, come with their specific cosmological constants $\hat{\la}$,  $\check{\la}$ and with their own energy momentum tensors. We make an assumption that both of them are the energy momentum tensors of mixtures of perfect fluids.
The Einstein equations for $\hat{g}$ are:
\beq
\hat{R}_{\mu\nu}-\frac12 \hat{R} \hat{g}_{\mu\nu}+\hat{\la} \hat{g}_{\mu\nu}=8\pi G \hat{T}_{\mu\nu},
\label{eqE}
\eeq
with:
\beq
\hat{T}_{\mu\nu}=\sum_i(\hat{\mu}_i+\hat{p}_i)\hat{u}_\mu \hat{u}_\nu+\hat{p}_i \hat{g}_{\mu\nu}
\label{eqT}
\eeq
in which
\beq
\hat{p}_i=w_i\hat{\mu}_i.
\label{perffluid}
\eeq
We also have the identically looking equations for $\check{g}$ with all the hats replaced by the checks. The respective velocities of the fluids are:
\beq
u_\mu=(1,0,0,0),\ \ \ \check{u}_\mu=\frac{1}{f} u_\mu\ \ \ \hat{u}_\mu=f u_\mu
\eeq

We assume that the cosmological constant is non-negative for both $\hat{g}$ and $\check{g}$. Consequently we write:
\beq
\check{\la}=3 \check{H}_{\la}^2,\ \ \ \ \hat{\la}=3 \hat{H}_{\la}^2
\eeq
In addition we assume that each fluid $(\hat{\mu}_i,\hat{p}_i)$ considered separately, satisfies its own continuity equation\footnote{This assumption is stronger than the continuity of the mixture of the fluids $(\sum_i\mu_i,\sum_ip_i)$. It is a particular solution of the continuity equation for the mixture assuming neither interaction among the fluids nor conversion of one fluid into another.}
\beq
\hat{\mu}_i'+3(\hat{\mu}_i+\hat{p}_i)\left(\ln\frac{h}{f}\right)'=0.
\eeq
An analogous assumption about the continuity of each fluid  $(\check{\mu}_i,\check{p}_i)$ leads to an analogous continuity equations for the checked quantities.
These solve as
\bea
\hat{\mu}&=&\sum_i\frac{3\hat{\mu}_{0i}}{8\pi G}\left(\frac{h}{f}\right)^{-3-3\hat{w}_i}\nn\\
\check{\mu}&=&\sum_i\frac{3\check{\mu}_{0i}}{8\pi G}(hf)^{-3-3\check{w}_i},
\label{mudens}
\eea
while the rest of the Einstein systems for the hat and the check metrics reduce to only two ODEs:
\begin{equation}
f^2(h/f)'^2-\sum_i\hat{\mu}_{0i}(h/f)^{-1-3\hat{w}_i}-\hat{H}_{\la}^2 (h/f)^2=-\ka\label{fhf1}
\end{equation}
for the hats, and
\begin{equation}
\frac{(fh)'^2}{f^2}-\sum_i\check{\mu}_{0i}(hf)^{-1-3\check{w}_i}-\check{H}_{\la}^2 (fh)^2=-\ka,
\label{fhf}
\end{equation}
for the checks.

According to our assumptions, when $t\to 0$, then $h(t)\to 1$ and $f(t)\to Ct$. Assuming smoothness of $f(t)$ and $h(t)$, i.e. assuming that their expansions in $t$ around $t=0$ have only integer powers of $t$, we note\footnote{It can also be seen by subtracting the two equations in (\ref{fhf}) from each other, which gives
\beq
\frac{(f^2)'(h^2)'}{f^2}+\sum_i\hat{\mu}_{0i}(fh)^{-1-3\hat{w}_i}+\hat{H}_{\la}^2 (fh)^2-\sum_i\check{\mu}_{0i}(h/f)^{-1-3\check{w}_i}-\check{H}_{\la}^2 (h/f)^2=0
\eeq} that to match the powers of $t$  when solving (\ref{fhf1})-(\ref{fhf}), we are forced to forbid majority of values for the constants $w_i$ (both hatted and checked). Actually we remain only with $w_i$s that are integer multiples of $\frac13$. Making the usual assumption that each $w$ satisfies $-1\le w\le 1$, gives us only \emph{four} possibilities (the biggest $w=\frac13$ corresponds to the the maximal inverse power of $t$ i.e. $1/t^2$, and the smallest one, $w=-1$, corresponds to the, already included separately, cosmological constant):
\beq
w_i=\frac{i}{3},\ \ i=-2,-1,0,1
\eeq
so $w$'s have to be ``quantized''.

The first two values -- with negative pressure -- are usually not considered in cosmology, but we should notice that $w=-1/3$ corresponds to a \emph{gas of strings}, while $w=-2/3$ corresponds to a \emph{gas of domain walls}. The two remaining values are usually the only ones used in cosmology on physical grounds: $i=1$ for \emph{radiation}, and $i=0$ for \emph{matter}. We find it interesting that these physically justified values of $w$, come out naturally from the requirement of the smoothness of conformal transformations relating the consecutive eons. In the following, to distinguish these two physical contributors to the fluid mixtures, we will use subscript $r$ (radiation) instead of $i=1$, $m$ (matter) instead of $i=0$, $d$ (domain walls) instead of $i=-2$ and $s$ (strings) instead of $i=-1$.

Introducing
\beq
\hat{F}=\frac{f}{h},\ \ \ \ \ \check{F}=fh
\eeq
we rewrite (\ref{fhf1})-(\ref{fhf}) as
\bea
h^2\hat{F}'^2-\hat{\mu}_{0r}-\hat{\mu}_{0m}\hat{F}-\hat{\mu}_{0d}\hat{F}^3-\hat{H}_{\la}^2 \hat{F}^4&=&(\hat{\mu}_{0s}-\ka) \hat{F}^2\nn\\
h^2\check{F}'^2-\check{\mu}_{0r}\check{F}^4-\check{\mu}_{0m}\check{F}^3-\check{\mu}_{0d}\check{F}
-\check{H}_{\la}^2&=&(\check{\mu}_{0s}-\ka)\check{F}^2
\label{FhF}
\eea

We have a relation coming from the leading $1/t^2$ divergence
\beq
C^2=\hat{H}_{\la}^2=\check{\mu}_{0r}
\eeq
It is tempting to assume that the both equations in (\ref{FhF}) are in fact one and the same equation i.e. that
\bea
\hat{\mu}_{0r}&=&\check{H}_{\la}^2,\ \ \ \ \check{\mu}_{0r}=\hat{H}_{\la}^2\nn\\
\hat{\mu}_{0m}&=&\check{\mu}_{0d},\ \ \ \ \check{\mu}_{0m}=\hat{\mu}_{0d}\nn\\
\hat{\mu}_{0s}&=&\check{\mu}_{0s}
\label{dualeq}
\eea
and then (with $h(0)=1$) we have
\beq
\hat{F}=\check{F}\ \Ra\ h(t) \equiv 1
\eeq
In such a case the \emph{intermediate metric} $g$, as in (\ref{gmetric}), is the metric of a \emph{static Einstein Universe}, and the solution to (\ref{FhF}) is given by the \emph{inverse elliptic functions}.
It is interesting to note that the radiation is dual to the cosmological constant, matter to a gas of domain walls, and a gas of strings is selfdual on two sides of the transition.

Assuming that $h(t)\equiv 1$ the Hubble parameter for $t\to 0$ is equal to
\beq
H(t)=- \frac{\rd}{\rd t}\left(\frac{1}{f}\right)\to \frac{1}{Ct^2}
\eeq
so the criticality, using (\ref{mudens}) and (\ref{dualeq}) tends to 1 for very early times:
\beq
\Om=\frac{8\pi G\hat{\mu}}{3H^2}\to 1.
\eeq
Thus in our model the Universe is critical at the early times. Note that it is also critical at the end when the dominant contribution to the energy comes from the cosmological constant.

\vspace{2mm}
Summarizing, we have shown the following:

Assume that
\begin{itemize}
\item the two consecutive eons with positive cosmological constants are filled with a mixture of perfect fluids, and
\item that their metrics are related, by an analytic conformal transformation, to the intermediate FLRW metric having constant spatial curvature,
\end{itemize}
then
\begin{itemize}
\item the fluids can only belong to 4 classes (radiation, dust and two other classes with negative pressures corresponding to a gas of strings and a gas of domain walls), and
\item the equations for the eons exhibit certain duality with the Einstein universe as the intermediate metric $g$, and
\item both eons (one at the end and the other at the beginning) are almost critical, so that the future eon dominated by radiation resembles the beginning of our Universe.
\end{itemize}

\vspace{0.2cm}
\noindent{\bf {Acknowledgments:}} We gratefully acknowledge helpful discussions with Roger Penrose and Paul Tod. K.A.M. was supported by the Polish NCN grant DEC-2013/11/B/ST2/04046, and P. N. by the Polish NCN grant DEC-2013/09/B/ST1/01799.


\begin{thebibliography}{99}




\bibitem{P} R. Penrose, {\it Cycles of Time: An Extraordinary New View of the Universe}, Bodley Head, 2010.

\bibitem{PT1} K.P. Tod,  Class. Quantum Grav. {\bf 20}, (2003) 521.

\bibitem{PT2} K.P. Tod, Gen.Rel.Grav. {\bf 47} (2015) no.3, 17.

\bibitem{ETN} E.T.Newman, {\it A Fundamental Solution to the CCC equation}, {\tt arXiv:1309.7271}


\end{thebibliography}
\end{document}